\documentclass{article}
\usepackage{arxiv}

\usepackage[utf8]{inputenc} 
\usepackage[T1]{fontenc}    
\usepackage{hyperref}       
\usepackage{url}            
\usepackage{booktabs}       
\usepackage{amsfonts}       
\usepackage{nicefrac}       
\usepackage{microtype}      
\usepackage{lipsum}
\usepackage{graphicx}
\usepackage{amsmath}
\usepackage{multirow}
\usepackage{svg}

\title{Performance of Objective Speech Quality Metrics on Languages Beyond Validation Data: A Study of Turkish and Korean}

\author{
  Javier Perez\thanks{Corresponding author.} \\
  Dept. of Intelligent Systems\\
  EEMCS faculty\\
  Delft University of Technology\\
  The Netherlands\\
  \texttt{javipeloza@gmail.com} \\
   \And
 Dimme de Groot \\
   Dept. of Intelligent Systems\\
   EEMCS faculty\\
  Delft University of Technology\\
  The Netherlands\\
  \texttt{d.c.c.j.degroot@tudelft.nl} \\
  \And
 Jorge Martinez\\
  Dept. of Intelligent Systems\\
  EEMCS faculty\\
  Delft University of Technology\\
  The Netherlands\\
  \texttt{j.a.martinezcastaneda@tudelft.nl} \\
}

\begin{document}
\maketitle
\begin{abstract}
Objective speech quality measures are widely used to assess the performance of video conferencing platforms and telecommunication systems. They predict human-rated speech quality and are crucial for assessing the systems quality of experience. Despite the widespread use, the quality measures are developed on a limited set of languages. This can be problematic since the performance on unseen languages is consequently not guaranteed or even studied. Here we raise awareness to this issue by investigating the performance of two objective speech quality measures (PESQ and ViSQOL) on Turkish and Korean. Using English as baseline, we show that Turkish samples have significantly higher ViSQOL scores and that for Turkish male speakers the correlation between PESQ and ViSQOL is highest. These results highlight the need to explore biases across metrics and to develop a labeled speech quality dataset with a variety of languages.
\end{abstract}


\section{Introduction}
Remote voice communication has become an important part of daily life. This can be through video conferencing platforms such as \textit{Zoom} and \textit{Teams}, but also over more traditional telecommunication systems. To ensure that users of these systems have an optimal experience, it is important to measure speech quality. The International Telecommunication Union (ITU-T) has established standardized metrics for measuring speech quality using both subjective and objective methods (e.g. \cite{itup563}, \cite{ITU800}, \cite{ITU808}). Subjective methods involve human ratings of speech signals, which makes them reflect the true opinion of users. However, to obtain accurate results these tests require a large number of listeners, making subjective tests expensive in both time and resources \cite{Shen2024}. To alleviate this problem, objective speech quality measures have been proposed as a more efficient and automated alternative to estimate speech quality. They are, however, less accurate than well-made subjective tests \cite{benefits_objective}.

Objective speech quality measures generally aim to predict the response a group of listeners would give in a subjective test. They typically achieve this by analyzing the degraded speech signal and mapping their internal representation to a human-interpretable MOS (mean opinion score) value ranging from 1 (``bad quality") to 5 (``excellent quality") \cite{Hines2015}. Objective measures can be categorized as intrusive and non-intrusive metrics. Intrusive metrics require a clean reference signal for comparison, estimating quality by measuring perceptual differences between the clean and degraded signals \cite{intrusive_metrics, full_reference}. Examples include PESQ \cite{pesq_original}, POLQA \cite{polqa}, and ViSQOL \cite{Hines2015, visqol}. On the other hand, non-intrusive metrics assess speech quality without a reference signal \cite{benefits_objective}. Prominent examples include ITU-T P.563 \cite{itup563} and deep-learning based models such as MOSnet \cite{Lo2019}, Quality-Net \cite{Fu2018}, NISQA \cite{Mittag2021} and DNSMOS \cite{Reddy2021}.

PESQ was standardized by the ITU-T in 2001 to predict subjective quality for narrowband telephony \cite{pesq_original}. PESQ processes reference and degraded signals, aligns them, and simulates human auditory perception \cite{pesq_diagram_ref}. The output is a score from -0.5 to 4.5, later mapped to a 1–5 MOS-LQO scale (Mean Opinion Score - Listening Quality Objective). The mapping, based on nine languages, achieves a precision of 93. 5\% within 0.5 MOS but lacks validation in other languages \cite{pesqmapping}. While PESQ was withdrawn in 2024 \cite{pesqwithdrawn}, it is highly likely that it will continue to be used in the research community and therefore remains of interest in this work. 

ViSQOL was developed by Google and the University of Dublin in 2015 and improves on PESQ for modern Voice over IP (VoIP) systems \cite{visqol}. ViSQOL converts speech into spectrograms, then compares similarity using a \textit{neurogram similarity index} (NSIM) and accounts for timing imprecision. The output is then mapped to a human-interpretable 1–5 MOS-LQO scale. ViSQOL performs well across various degradation types but relies on unpublished subjective datasets \cite{visqol_performance}.

While the use of these objective metrics is widespread, a major limitation in their design is that their validation and development relies on a limited set of languages. For example, PESQ and its successor POLQA (Perceptual Objective Listening Quality Analysis) \cite{polqa}, are calibrated using data from nine and ten languages \cite{pesqmapping} respectively. This limited set of languages raises concerns about their accuracy when applied to other languages. Given the growth of multilingual communication and applications, analyzing and evaluating the robustness and accuracy of objective speech quality metrics across diverse languages is important. 

In this paper we evaluate PESQ and ViSQOL for Turkish and Korean, comparing them to English as a reference. Turkish and Korean are chosen due to their significant internet penetration rate (86.5\% and 97.2\% respectively \cite{internet_languages}). We explore how these metrics perform in predicting speech quality for these languages, further considering gender, and we test for different degradation types. Prior studies have examined PESQ and POLQA in African and Arabic languages \cite{benali_language_dependence_2002, Konane2021Impact}, highlighting variations in performance. This study expands on these findings by assessing PESQ and ViSQOL’s performance on languages not in their original mapping function validation set.

\section{Methodology}
This section outlines the methodology for evaluating speech quality across languages under various degradation conditions. Reference samples are degraded using different noise types and signal-to-noise ratios and the resulting signals are used to assess objective quality metrics.

\subsection{Dataset}
We use the open-source ALLSSTAR Corpus Multilingual Dataset \cite{ALLSSTAR}. This corpus features men and women native speakers reading text in multiple languages. The languages in this dataset include among others English, Turkish, and Korean, recorded under equal conditions. The speech samples are recorded in high-quality PCM 16-bit format with a bit rate of 353 kbps and a 22.05 kHz sampling rate and are considered clean for the purposes of this study.

For each language, 16 continuous speech samples are extracted, with an equal distribution of 8 men and 8 women speakers to minimize gender bias. The speakers ages are between 18 years and 29 years. The average age is 22 years. Each speech sample lasts between 5 and 10 seconds. We minimize the amount of silence in the excerpts.

\subsection{Objective Quality Measures}
We use PESQ and ViSQOL to estimate the objective speech quality. For PESQ, we use the open-source Python implementation by Wang et al. \cite{pesq_implementation}. For ViSQOL we use the implementation in the Audio Toolbox in MATLAB R2024b \cite{visqolmatlab}. ViSQOL is used in \texttt{speech} mode, which requires the input signals to be band-limited to 16 kHz.

\subsection{Signal Degradation Types}
Various types of noise are added to the speech signal in order to simulate different real-world conditions and assess the performance of the quality metrics in these cases. These synthetic degradations are commonly used to replicate environments where speech quality may be affected by background interference \cite{types_of_noise}:

\begin{itemize}
    \item \textbf{Pink Noise:} With power density inversely proportional to frequency, it mimics natural sounds (e.g. rain, wind) and is chosen for its common occurrence in outdoor environments \cite{blue_noise}.
    \item \textbf{Blue Noise:} With power density proportional to frequency, it has the opposite distribution of that of pink noise \cite{blue_noise}.
    \item \textbf{Babble Noise:} Sounds that resemble overlapping human voices. It simulates noisy crowded environments, like restaurants \cite{babble_noise_explained}.
\end{itemize}

\subsection{Reference Signal and Degraded Signal Generation}
In order to simulate a classical telecommunication channel, the reference signal and the degraded signal used for the evaluation are created by applying the following processes to the original speech samples:


\begin{enumerate}

\item \textbf{Down-sampling and IRS Filtering}: Speech and noise are mixed into a monaural channel and down-sampled to 8 kHz, simulating bandwidth limitations of legacy telecommunication systems. Afterwards, the signals are filtered using an Intermediate Reference System (IRS) filter restricting the bandwidth to 0.3-3.4 kHz to create a narrowband signal \cite{itu_t_rec_p48}.

\item \textbf{Normalization}: Speech and noise signals are normalized to -26 dBFS (decibels relative to full scale) to ensure consistent levels across all samples and enough headroom to prevent clipping. The speech signal obtained during this step serves as the reference signal in the experiment.

\item \textbf{Mixing}: Noise is added to the speech signals at varying Signal-to-Noise-Ratio (SNR) levels from -25 dB to 40 dB. The equations for calculating the required gain to achieve the correct SNR for the noise signal are given in (\ref{eq:rms_noise}) and (\ref{eq:gain}), 
    \begin{equation}
        P_d = \frac{P_s}{10^{T/20}}
        \label{eq:rms_noise}
    \end{equation}
where $P_d$ is the desired Root Mean Square (RMS) noise power, $P_s$ is the RMS power of the clean signal, $T$ is the target SNR in dB, and,    
    \begin{equation}
        G = 20 \log_{10} \left( \frac{P_d}{P_n} \right)
        \label{eq:gain}
    \end{equation}
where $G$ is the desired gain, and $P_n$ is the current RMS noise power.
\item \textbf{Re-normalization}: After mixing, the combined signal is re-normalized to -26 dBFS for consistent loudness.

\item \textbf{Encoding/Decoding}: The final signal is encoded and decoded using the G.711 A-law codec \cite{itu_t_g711_0, ti_alaw_mulaw} to simulate the effects of compression and decompression on speech quality. This is often used in telecommunications systems to minimize the perceptibility of the quantization noise produced by 8 bit quantization

\end{enumerate}

The final dataset consists of 2016 degraded signals: \(14\, \text{ SNR levels} \times 3 \, \text{ degradation types} \times 16 \, \text{ samples per language} \times 3 \, \text{ languages}\). These degraded signals are compared to a total of 48 reference signals: \(16 \text{ samples per language} \times 3 \text{ languages}\).

\section{Results} \label{sec:results}

This section presents and discusses the results obtained from conducting the previously described experiment of evaluating PESQ and ViSQOL against Turkish, Korean and English degraded samples.

\begin{figure}[ht]
    \centering
    \includegraphics[width=0.6\textwidth]{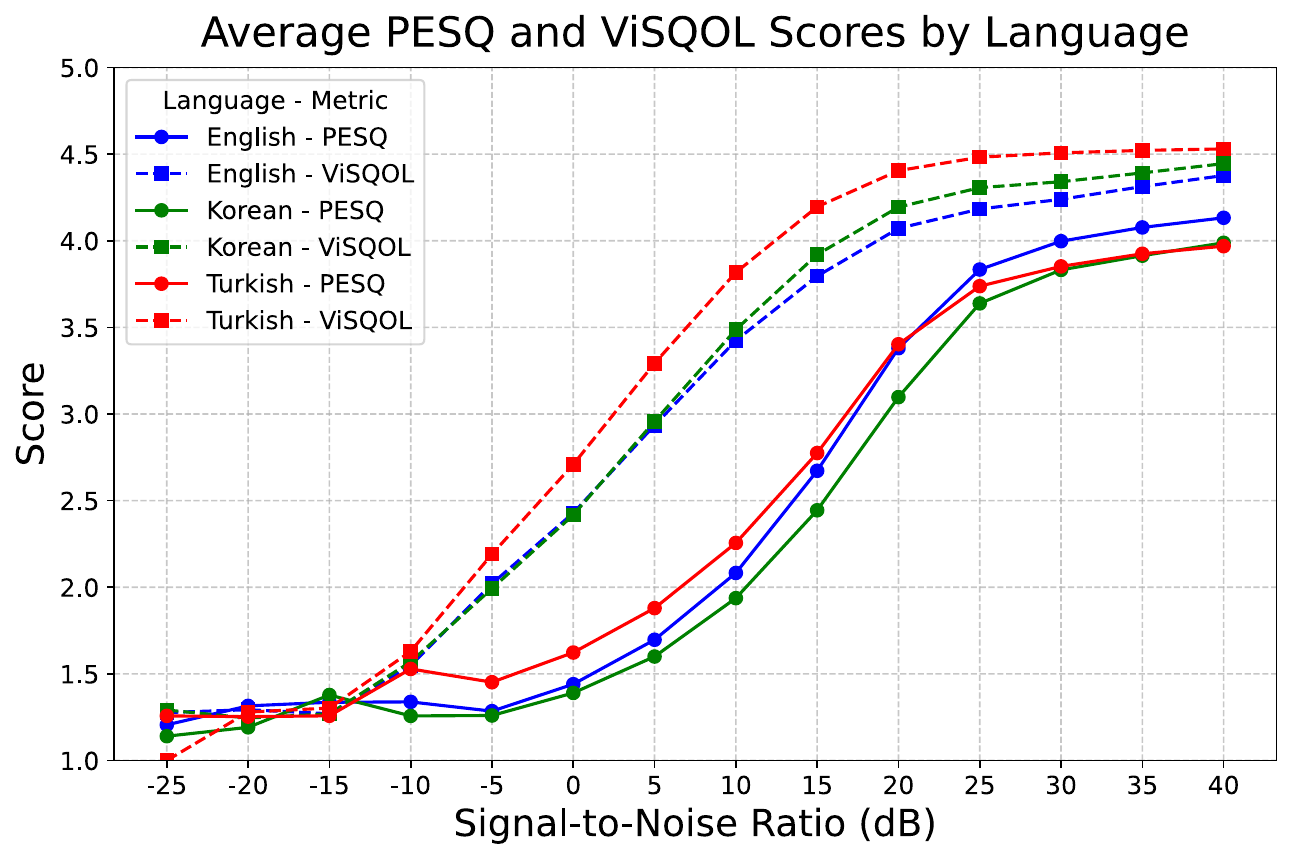}
    \caption{Average PESQ and ViSQOL MOS-mapped scores (1 = `bad', 5 = `excellent') across all degradation types, segmented by language, with SNR values from -25 dB to 40 dB.}
    \label{fig:comparison_pesq_visqol_by_snr}
\end{figure}

Figure \ref{fig:comparison_pesq_visqol_by_snr} shows the average MOS-mapped scores for PESQ and ViSQOL across different SNR values and languages. The scores returned by the algorithms are only the final mapped scores, which are obtained after applying the mapping function. These results reveal no significant difference between the PESQ scores across languages. However, ViSQOL shows a consistent trend, with Turkish scores averaging 4.95\% higher than English scores, corresponding to a 0.19 point difference. This difference increased to 10.18\% (0.34 points) within the SNR range of 0 dB to 25 dB.

\begin{figure}[ht]
    \centering
    \includegraphics[width=0.7\textwidth]{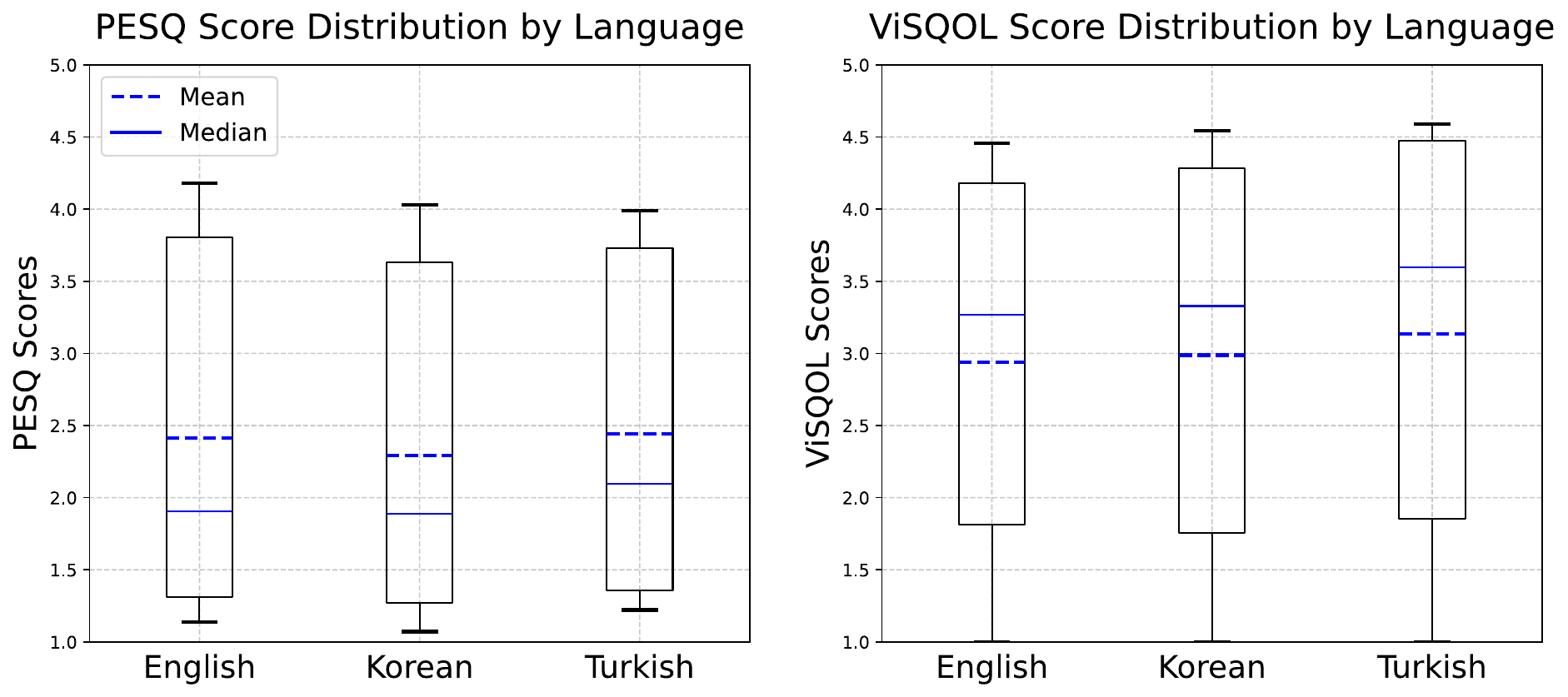}
    \caption{Boxplots of PESQ and ViSQOL scores (1 = `bad', 5 = `excellent') by language, showing the distribution of speech quality metrics.}
    \label{fig:boxplots_by_metric}
\end{figure}

In Figure \ref{fig:boxplots_by_metric}, key statistical features of PESQ and ViSQOL score distributions across languages are further shown. Here it can be seen the median ViSQOL score for Turkish is 9\% higher than for English and Korean.

\begin{figure}[h]
     \centering
     \includegraphics[width=0.6\textwidth]{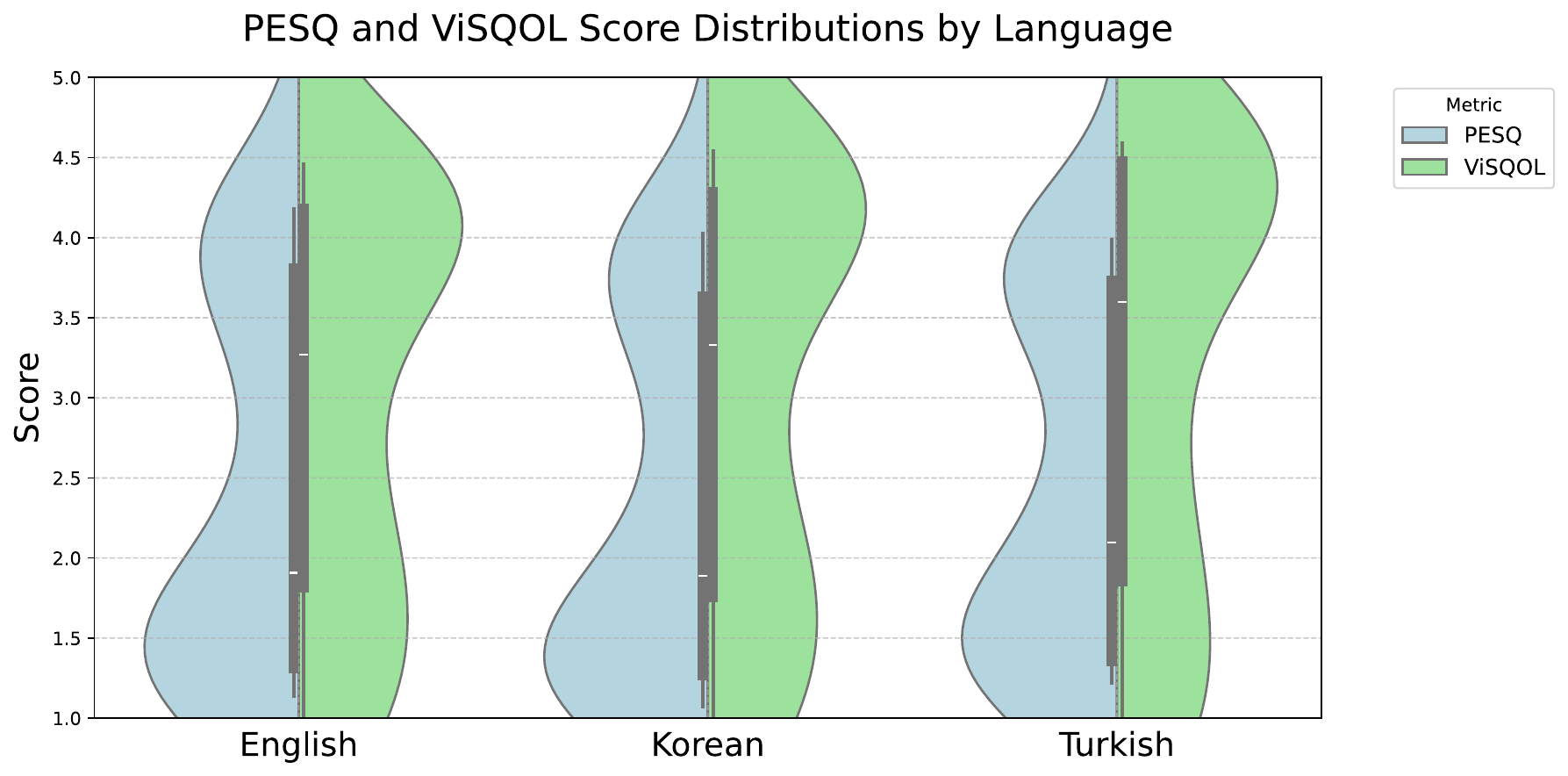}
     \caption{Density distributions of PESQ (blue) and ViSQOL (green) scores (ranging from 1 = `bad', to 5 = `excellent' quality) by language, illustrating the variation in score distributions across languages.}
     \label{fig:score_density}
 \end{figure}

This information alone is insufficient to determine whether the differences in Turkish scores are statistically significant compared to the rest. To further assess the impact of language on objective speech quality ratings, we used the Kolmogorov-Smirnov (KS) test \cite{kstest}, a non-parametric method of which the null hypothesis is that two independent samples (PESQ and ViSQOL scores for different languages in our case) come from the same distribution. The KS test is chosen due to the bimodal nature of the distributions (as can be seen in Figure \ref{fig:score_density}), making traditional parametric tests like t-tests or ANOVA unsuitable. These tests assume normal, unimodal distributions, which does not apply here, while the KS test is flexible and does not require such assumptions \cite{statistical_metrics}.

A significance threshold of the resulting p-value is set to 0.05 for this study. This value is a commonly chosen to balance the risk of Type I errors (false positives) while maintaining practical significance \cite{significance_threshold, pvalue_opinion}. 

Table \ref{tab:ks_test_results} shows the KS test results for PESQ and ViSQOL scores. For PESQ, all pairwise comparisons had p-values greater than 0.05, meaning the null hypothesis cannot be rejected, and language did not significantly affect PESQ scores. For ViSQOL, a significant difference is found between English and Turkish, with a p-value of 0.02, indicating a significant difference in their distributions. However, the other pairwise comparisons have p-values greater than 0.05 suggesting no significant differences in those cases. While failing to reject the null hypothesis does not prove no differences exist, it implies any observed differences could be due to random variation, and further investigations may be needed to arrive to stronger conclusions.

\begin{table}[th]
  \caption{Kolmogorov-Smirnov (KS) test results for PESQ and ViSQOL scores segmented by language.}
  \label{tab:ks_test_results}
  \centering
  \begin{tabular}{c c c c}
    \toprule
    \textbf{Metric} & \textbf{Comparison} & \textbf{KS-statistic} & \textbf{p-value} \\
    \midrule
    \multirow{3}{*}{PESQ}  
      & English vs Korean   & 0.17 & 0.61 \\
      & English vs Turkish  & 0.20 & 0.44 \\
      & Korean vs Turkish   & 0.21 & 0.29 \\
    \midrule
    \multirow{3}{*}{ViSQOL} 
      & English vs Korean   & 0.14 & 0.79 \\
      & English vs Turkish  & 0.33 & \textbf{0.02} \\
      & Korean vs Turkish   & 0.23 & 0.18 \\
    \bottomrule
  \end{tabular}
\end{table}

Overall, it is found that PESQ scores tend to be lower, while ViSQOL scores are higher. This is shown in Figure \ref{fig:boxplots_by_metric}, where the PESQ boxplot shows the median below the mean for all languages, indicating a concentration of lower scores, and the ViSQOL boxplot has the median above the mean, suggesting higher values are more concentrated. 

Figure \ref{fig:score_density} illustrates with a violin plot the density distribution of PESQ and ViSQOL scores. PESQ scores peak at an average of 1.44, concentrated at the lower end with a longer right tail. In comparison, ViSQOL scores peak at 4.19, clustered at the higher end with a longer left tail.

\begin{figure}[h]
    \centering
    \includegraphics[width=0.6\textwidth]{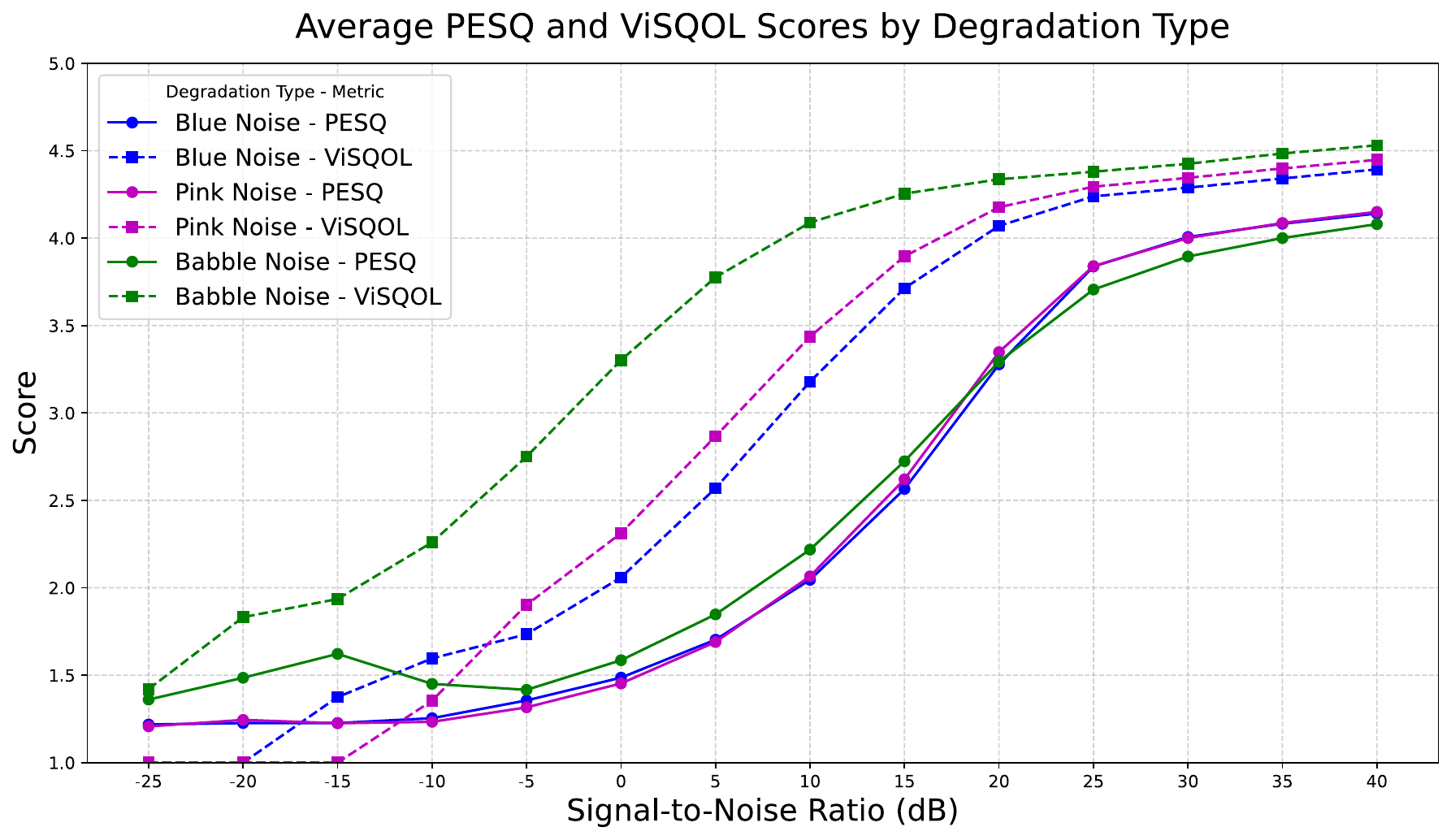}
    \caption{Evolution of PESQ and ViSQOL scores (ranging from 1 = `bad', to 5 = `excellent' quality) segmented by degradation type with SNR values from -25 dB to 40 dB.}
    \label{fig:metrics_by_noise_type}
\end{figure}

Figure \ref{fig:metrics_by_noise_type} compares the average PESQ and ViSQOL scores across different degradation types. It further confirms that ViSQOL scores are generally higher than PESQ scores, with the largest difference observed in samples affected by babble noise. The average difference for babble noise is 72\% greater than that for pink and blue noise, indicating that crowd noise has a smaller impact on ViSQOL scores for the languages tested.

Table \ref{tab:ks_test_languages} presents the KS test results for PESQ and ViSQOL scores across different degradation types. It shows that babble noise comparisons in ViSQOL return the lowest p-values, with a significant difference between babble and blue noise (p-value = 0.04). The p-value for the pink and babble noise comparison is 0.06 suggesting significance, but further investigation with more samples and noise types is recommended to have stronger confirmation.

\begin{table}[th]
  \caption{Kolmogorov-Smirnov (KS) test results for PESQ and ViSQOL scores segmented by degradation type.}
  \label{tab:ks_test_languages}
  \centering
  \begin{tabular}{c c c c}
    \toprule
    \textbf{Metric} & \textbf{Comparison} & \textbf{KS-statistic} & \textbf{p-value} \\
    \midrule
    \multirow{3}{*}{PESQ}  
      & Blue vs Pink Noise  & 0.12 & 0.93 \\
      & Blue vs Babble Noise & 0.17 & 0.61 \\
      & Pink vs Babble Noise & 0.19 & 0.44 \\
    \midrule
    \multirow{3}{*}{ViSQOL} 
      & Blue vs Pink Noise  & 0.10 & 0.99 \\
      & Blue vs Babble Noise & 0.31 & \textbf{0.04} \\
      & Pink vs Babble Noise & 0.29 & 0.06 \\
    \bottomrule
  \end{tabular}
\end{table}

\begin{figure}[ht]
    \centering
    \includegraphics[width=0.6\textwidth]{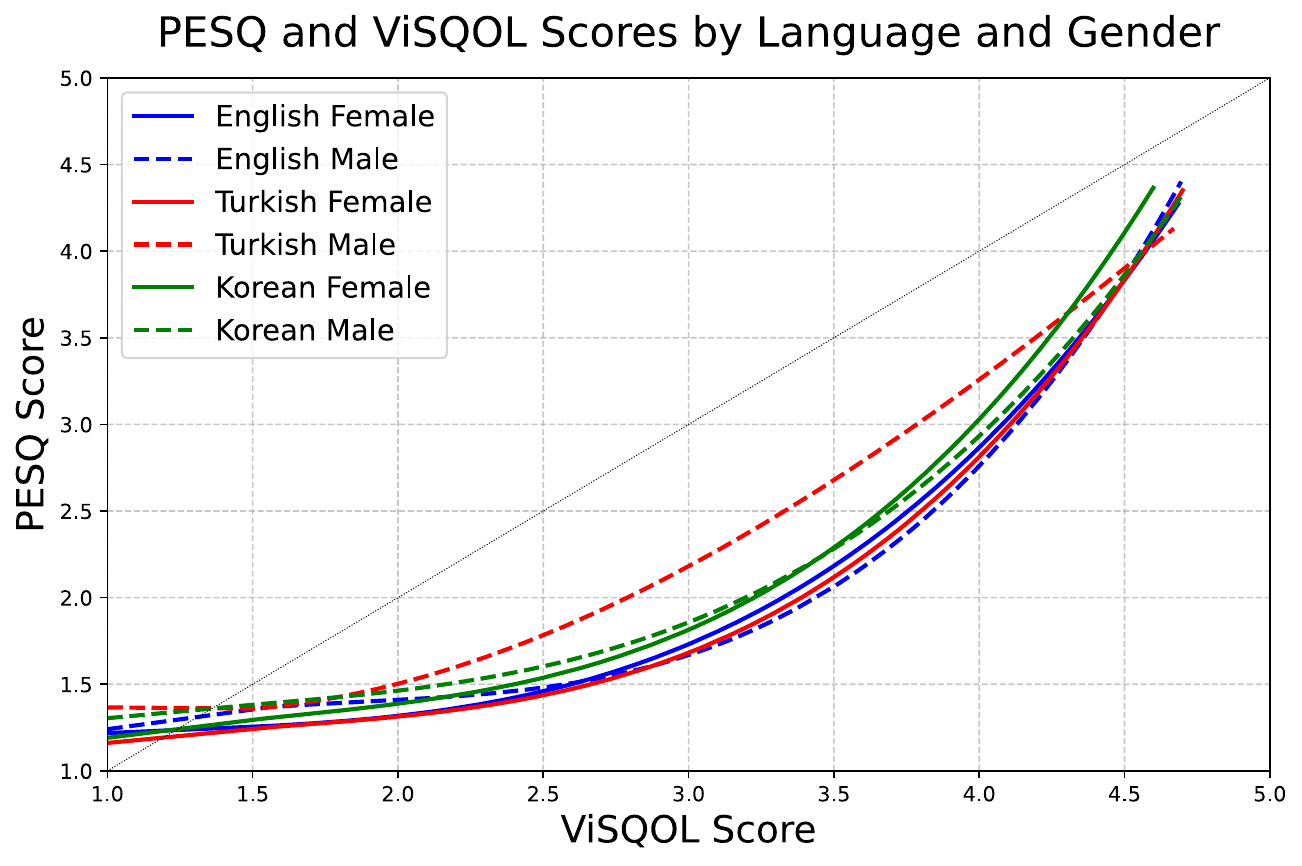}
    \caption{PESQ and ViSQOL scores (ranging from 1 = `bad', to 5 = `excellent' quality) correlation segmented by gender, with data fitted using a cubic polynomial function.}
    \label{fig:pesq_visqol_correlation_gender}
\end{figure}

To analyze the impact of gender, Figure \ref{fig:pesq_visqol_correlation_gender} shows the correlation between PESQ and ViSQOL scores by language and gender. While all language groups follow a similar trend, Turkish men show a distinct pattern with a smaller gap between PESQ and ViSQOL scores. This was observed after fitting a cubic polynomial function to the scores.

For a more detailed analysis, Table \ref{tab:statistics_correlation} provides a statistical comparison of PESQ and ViSQOL scores across languages and genders. It includes metrics like mean absolute deviation (MAD), root mean squared deviation (RMSD), and mean differences in scores. Results are presented for the overall dataset, a group excluding Turkish male speakers (Non-TM), and Turkish male speakers (TM). The final column highlights the differences between Turkish male speakers and the non-Turkish male group to reveal any specific variations.

\begin{table}[ht]
  \caption{Statistical metrics: mean average deviation (MAD), root mean squared deviation (RMSD), and mean difference, comparing the correlation of PESQ and ViSQOL scores for the overall dataset, average scores excluding non-Turkish male speakers (Non-TM), Turkish male speakers (TM), and the difference between the last two groups (Diff).}
  \label{tab:statistics_correlation}
  \centering
  \begin{tabular}{c c c c c}
    \toprule
    \textbf{Metric} & \textbf{Overall} & \textbf{Non-TM} & \textbf{TM} & \textbf{Diff} \\ 
    \midrule
    MAD  & 0.71 & 0.73 & 0.62 & -0.11 \\ 
    RMSD & 0.89 & 0.91 & 0.77 & -0.13 \\ 
    Mean difference & -0.62 & -0.65 & -0.47 & 0.18 \\ 
    \bottomrule
  \end{tabular}
\end{table}

As seen, Turkish male speakers show lower MAD and RMSD values compared to both the overall dataset and the non-Turkish male group. Specifically, their MAD is 0.62, which is 17.25\% lower than the non-Turkish male group, and their RMSD is 0.77, 17.38\% lower. These results suggest that the alignment between PESQ and ViSQOL scores is more consistent for Turkish male speakers.

Additionally, the mean difference between PESQ and ViSQOL scores is 0.18 lower for Turkish male speakers, representing a 37.9\% higher alignment compared to non-Turkish males. This indicates that, on average, PESQ and ViSQOL scores are more closely aligned for Turkish male speakers, while non-Turkish males show a larger discrepancy between the two metrics.

\section{Discussion} \label{sec:research_limitations}
A limitation of this research is that the dataset used has a limited number of speakers per language. Therefore, to achieve an equal man-to-woman ratio, repeated speakers were added in some cases. Additionally, the speakers ages ranged only from 18 to 29 years old.

The study assumes uniform recording quality across languages. While the degradation types used are representative approximations, they do not cover all real-world scenarios. Lastly, the lack of datasets with subjective MOS ratings for the analyzed languages means only a statistical analysis was possible. A dataset including these subjective ratings would provide a more reliable test.

\section{Conclusions and future work} \label{sec:conclusion}
This paper assesses PESQ and ViSQOL for predicting speech quality in Turkish and Korean, two languages outside the mapping function validation sets of these objective quality measures. While results are similar to English, Turkish showed higher ViSQOL scores, with a 5\% average increase, reaching 10\% at mid-range SNR values. Kolmogorov-Smirnov tests reveal a significant difference only between English and Turkish ViSQOL scores (p = 0.02).

Furthermore, ViSQOL scores are more concentrated at the upper end (higher quality), while PESQ scores are generally lower. Babble noise shows the largest discrepancy between the two metrics, suggesting ViSQOL is less sensitive to background speech-like noise. Kolmogorov-Smirnov (KS) tests confirm significant differences in ViSQOL scores between blue and babble noise (p = 0.04), and a p-value of 0.06 for pink and babble noise, which are worth investigating in the future.

Lastly, gender analysis reveals Turkish men have the smallest gap between PESQ and ViSQOL scores, indicating a more consistent evaluation of speech quality in this group, with 37.9\% better alignment compared to others.

Future research might address the limitations discussed in subsection \ref{sec:research_limitations} and explore additional metrics such as POLQA and DNSMOS. Expanding the study to include more languages, age groups, degradation types (e.g. reverberation), and wideband audio would further demonstrate how general are these metrics in practice. Repeating the experiment on other languages from the PESQ and ViSQOL validation set would further help strengthen the results.

Determining the generalization of objective speech quality metrics across languages requires having sufficient labeled data with subjective quality scores. Building such a dataset would help identify biases, enable adjustments, and validate mapping functions, advancing future research.

\bibliographystyle{unsrt}  
\bibliography{references}  

\end{document}